# Self-Organized Networks and Lattice Effects in High Temperature Superconductors III: Dopant Internal Structure

J. C. Phillips

Dept. of Physics and Astronomy, Rutgers University, Piscataway, N. J., 08854-8019

## Abstract

The self-organized dopant percolative filamentary model, entirely orbital in character (no fictive spins), predicted the existence and even the ~ 3 nm diameters of gap nanodomains (discovered by STM) that amazed almost all theorists in this field. Here it explores the self-consistent and self-limiting coupling between dopant sites and gap character. It explains precursive magnetic effects and shows that they are caused by dopant ordering into loops at an onset temperature of order $2T_c$. The ordered coupling of these loops by filamentary connectors is reflected by Debye-Waller factors, and it apparently leads to HTSC. The model also resolves the mystery of the vanishing isotope effect, and it explains the origin of the high-energy "waterfall" recently discovered by ARPES, as well as some overlooked infrared anomalies.

## 1. Introduction

High temperature cuprate superconductivity may be the most complex phenomenon known in inorganic materials. It has been the subject of more than 65,000 papers, and a large number of theoretical models have attempted to explain the many counter-intuitive phenomena observed. Many of the theoretical papers contain elaborate and ingenious mathematical models whose relation to experiment is vague. In this paper topological methods, based on the author's earlier theories of the glassy behavior of dopants in the cuprates [1], are used to discuss several anomalies in detail, with emphasis on the key role played by the connectivity of the internal dopant structure. The theory emphasizes qualitative trends, as experience has shown that that the complexity of these materials may well preclude quantitative treatments of the kind that worked so well for simpler superconductors, such as $MgB_2$.

This paper is the third and last part of a longer paper that has been broken into parts to make it more digestible. (Certainly it is much more accessible than the 65,000 papers that have been written on the cuprates!) I have retained the Section, Figure and Reference numbering of the longer paper to avoid translational errors. The first part of this paper appeared as cond-mat/0611089 (I), and the second part as cond-mat/0611186 (II). This three-part presentation has been revised to include some new material in this rapidly evolving field.

**9. Nanodomains and Dopant Sites**

Wide-area STM experiments on $Bi_2Sr_2CaCu_2O_{8+x}$ (BSCCO) [5] have opened a unique window on the nanodomain gap structure, which has surprised almost all the theorists committed to models of HTSC based on homogeneous lattices. Fortunately for theorists, the geometry of STM studies of micaceous surfaces still leaves many questions unanswered. The internal gap structure near the surface is projected on a single plane, so that even if one assumes that only the uppermost unit cell dominates the observations, the gap patterns are still superpositions of patterns from at least two $CuO_2$ planes, and probably two BiO planes as well. The gaps are associated with off-lattice $O_x$ interstitials, which are not easily observed, and the role of these interstitials is not easily established.

Before one can suggest a theoretical model for dopant siting, one should first consider the main features of the observed patchy nanodomain structure. The patches are typically d ~ 3 nm in diameter, and this length is independent of dopant concentration from under- to over-doped [5]. It seems likely that the sizes of the patches are set by the self-limiting effects of internal stress, as these often limit cluster sizes in glasses (boson peak [59]) Because there are two kinds of gaps, it seems reasonable to assume that there are two kinds of interstitial sites – one associated with pseudogap nanodomains, and one associated with superconductive gap nanodomains.

McElroy *et al.* [60] have recently reported a remarkable observation: they have succeeded in identifying one of the two possible oxygen interstitial dopant sites, labeled *A* and *B*. They were able to localize the *A* dopants using enhanced currents in their I-V characteristics at -0.9 V below the Fermi energy $E_F$. The *A* dopant sites (located laterally with a resolution better than 1A) are positively correlated with large values of the strongly inhomogeneous gap parameter $\Delta(\mathbf{r})$ and are presumably associated with the pseudogaps, generally larger than superconductive gaps. The *B* dopant states of the latter should pin $E_F$, as they do in semiconductor impurity bands, and not be 1 eV below it, where they are part of the electrically inactive part host framework. Almost all of the STM data have been taken at low T, but recent data taken near $T_c$ have shown a very interesting effect: increasing T by 14% across $T_c$ produced no measurable change in the residual pseudogap distribution in an optimally doped BSCCO sample, but greatly enhanced $N(E_F)$ [61]. This is a possible signature of *B* dopants.

## 10. Precursive Magnetic Effects

One of the striking anomalies in HTSC is the presence of two very large precursive magnetic effects: the giant Nernst effect in the thermomagnetic power [62], and field-enhanced diamagnetism, with composition dependence parallel to that of the Nernst effect [63]. To understand these effects, let us look at hierarchical trends in possible sets of dopant configurations, assuming that the dopants minimize the free energy by optimally screening internal ionic fields.

The first point is that between the pseudogap nanodomains decorated by *A* dopants, there will be channels occupied by *B* dopants. In other words, the pseudogap regions represent barriers in a maze around which filamentary paths centered on electrically active dopants must percolate. Thus the *B* dopants are not randomly distributed (Fig. 8(a)). Instead, as with all disordered percolative structures, local structure is characterized by its longitudinal and transverse features (Fig. 8(b)). In the conducting regime there are local curvilinear channels that are longitudinally open, but transversely confined, with a variable spacing between the pseudogap confinement walls. Within the tetragonal framework of conductive $CuO_2$ planes one can distinguish two principal directions, the (10) ones parallel to CuO rows, and the (11), angularly between the (10) and (01) CuO rows. Electron-phonon scattering is strongest parallel to CuO rows, so one expects that mean free paths will be longer in the (11) directions than in the (10) directions. In other words, the Fourier grid of the host lattice lifts the planar angular degeneracy of the glassy dopant network.

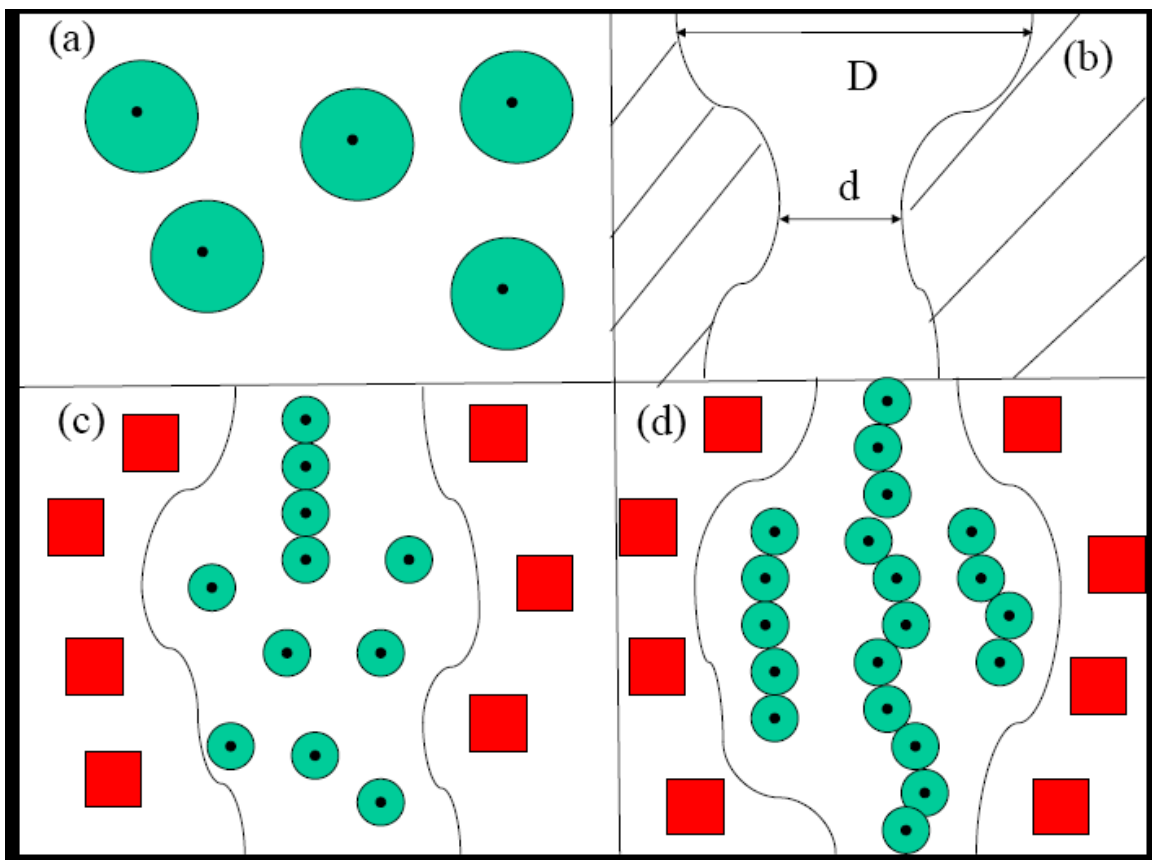

*Fig. 8. (a) Randomly distributed dopants (black dots) in a homogeneous background. Ionic fluctuations are screened dielectrically in regions here idealized as green circles. (b) Channel geometry formed by pseudogap regions, with variable widths typically ranging from d to D. (c) and (d); the black dots are $O_x^-$, while the red rectangles are the electrically inactive $O_x^{2-}$, and the green circles are the electrically active $O_x^-$. (c) Even at low doping densities it is energetically advantageous for the electrically active dopants (green circles) to form linear segments (upper chain). (d) More than one linear segment*

*can form in wide regions of a channel; when the segments lengthen, narrow regions act as bottlenecks, limiting the number of percolating segments.*

There are several factors that optimize dielectric screening. The screening is approximately exponential (classically, Debye-Waller; quantum-mechanically, Thomas-Fermi) around each dopant, but in our figures for simplicity this sceening charge is represented by a constant charge density with a radial cutoff (a disc in plan view). The actual "fuzzy" screening charge spreads out and can overlap with nearest neighbors. The overlap is favorable for screening because there is metallic charge delocalization, and higher conductivity along filamentary paths. However, if the charges overlap too much, then the volumes (or areas in plan view) are duplicated, and some screening energy is lost to redundancy. This factor tends to favor more nearly linear metallic filaments.

Another factor is the variability of channel widths. Suppose we call the wide channel regions ponds, and the narrow ones bottlenecks. *B* dopants will gain more energy from screening ponds than from screening bottlenecks, as confinement in the bottlenecks means that some of their screening charge will spread out into pseudogap regions, where the lattice is stiffened by the pseudogap, and hence the marginal instability of ionic fluctuation amplitudes is reduced. Because of the exponential fuzziness of the screening charge, they will also gain more energy from screening larger ponds than from screening smaller ones, especially as the former are well suited to the formation of dopant loops that can carry large screening ring currents, as the latter are not scattered by filamentary ends.

Another characteristic feature of ponds and bottlenecks is that they will generate loops of dopant centers in the ponds, as well as strings or filaments of dopant centers that extend from pond to pond through a bottleneck. Which feature is favored depends on details of channel geometries, the relative magnitudes of end energies, and the linearity energy. See Fig. 9 for examples of some of the possibilities.

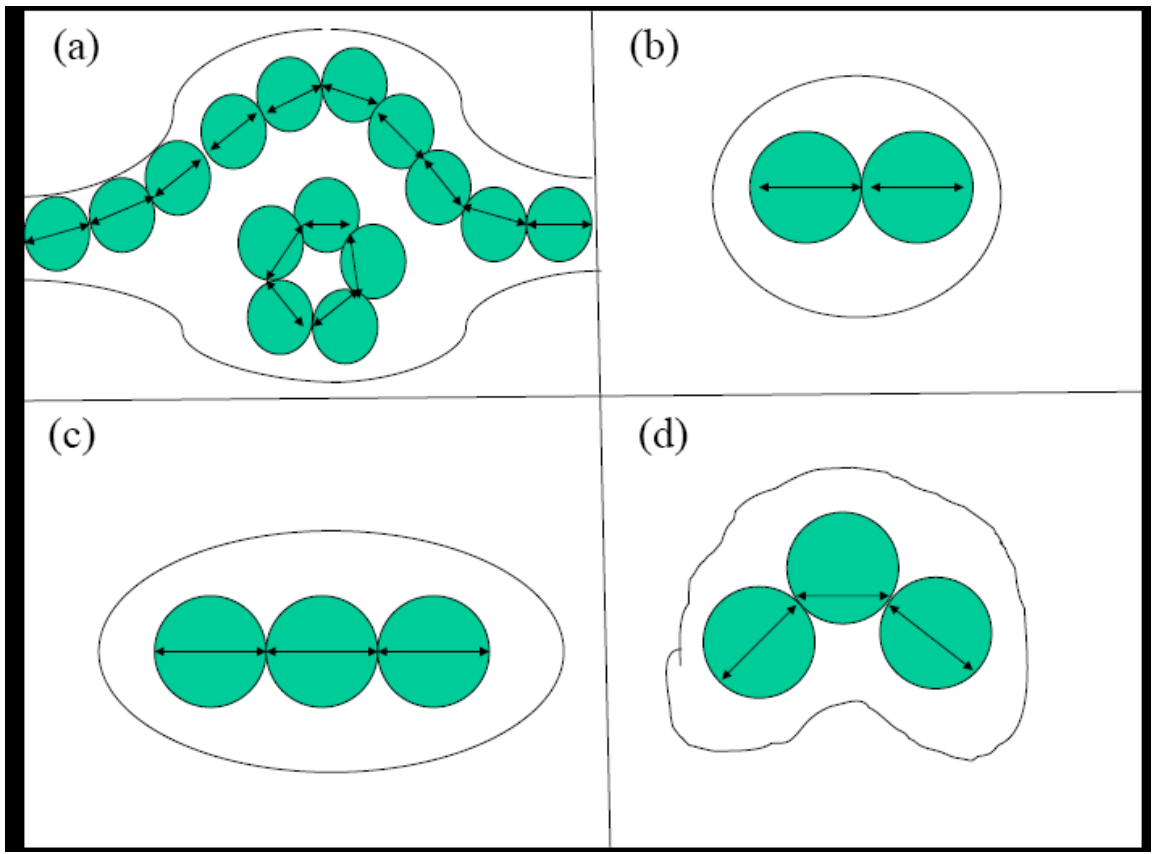

*Fig. 9. (a) When there is free area in wide regions which are shorter than shown in Fig. 1(d), closed loops (hydrodynamic eddies) are energetically favored at high temperatures. (b) Adjacent non-overlapping dopant pairs screen larger areas. Linear segments (c) screen larger areas than curved segments (d). In these figures, double arrows in each circle represent enhancement of screening along enhanced polarizability axes due to charge overlap (not shown).*

Each of these topological features has its own length scale, and with each length scale some kind of mean-field transition occurs. These cascading mean-field transitions eventually led to superconductivity, but they also correspond to hierarchies of stress-induced and strain limited transitions of the Martensitic type. The existence of such multiple length scales is a quite general consequence of off-lattice relaxation, and Bishop and coworkers have discussed it frequently [64]. Because the cuprates are spectacularly soft, with $R \approx 2$ (Fig. 2 of I), the effects here are larger and better-defined than in other soft lattices, for instance, the traditional ferroelectric and ferroelastic perovskites (such as BaTiO$_3$) where R ≈ 2.4.

Let us start with the highest temperatures, where filamentary loops begin to form. The virtual diamagnetic currents associated with isolated loops can be detected by spin-polarized neutron scattering, and such scattering has been observed [65]. The formation temperature T$_{mag}$ in YBCO$_{6 + x}$ ranges from 300K at x = 6.5 to 170K at x = 6.75, so that T$_{mag}$ is similar to a pseudogap transition temperature obtained from linearity of the planar resistivity. The mechanism proposed here, involving dopant loops, is different from the one suggested in [65], where only the host lattice is discussed, and translational invariance is supposed to be required. However, the dopant spacings in loops will also be regular, as these must be in registry with the lattice, and this weaker condition of registry is enough to cause the spin-polarized neutron scattering to peak at Bragg peaks. The signals produced by registry are weaker than those produced by translational invariance, which explains why the observed signals are also very weak. Moreover, the present model explains why T$_{mag}$ is similar to a pseudogap transition temperature, as the loops will begin to form as soon as the channels and ponds form. There is no obvious connection between T$_{mag}$ and a resistivity pseudogap transition temperature in the translationally invariant model.

It is noteworthy that the temperature dependence of the spin-polarized neutron scattering does not have the mean-field form f(1- (T$_{mag}$/T)$^2$) expected from translational invariance, i. e., microscopic homogeneity. Instead the temperature dependence is ragged, and more nearly linear in T$_{mag}$/T. This is consistent with the kind of broadening one would expect from a glassy model. As T is lowered, adjacent ponds will be filled more often, and strings will appear connecting such ponds. In the hydrodynamic analogy, eddies will appear in adjacent ponds, and streaming will carry particles from one eddy to the next.

Similarly, in the presence of crossed electric (or thermal gradient) and magnetic fields, the vorticity of carriers in these loops can be activated to jump across the bottlenecks to loops in adjacent "ponds". Because the hopping distance is large, and the pinning of the vortices by the soft host lattice is weak, this mechanism will inevitably lead to a giant Nernst effect in the thermomagnetic power. This effect has been observed [66] in both BSCCO and LSCO, and it peaks near x = 0.11, in other words, well above the MIT at x = $x_c$ = 0.06, yet well below optimal doping x = $x_0$ = 0.16, near x = ($x_c$ + $x_0$)/2, which would be the expected value for a plausible channel aspect ratio of 2. The onset temperatures at optimal doping $x_0$ are 150K (BSCCO) and 90K (LSCO), well above $T_c$, and well below pseudogap temperatures (and presumably $T_{mag}$, which is likely to be especially large in YBCO because the chains facilitate vortex hopping).

From the "tilted hill" magnetic field dependence that is observed both above and below $T_c$, the authors conclude [66] that the vortices are continuously present in the normal as well as the superconductive state. This agrees perfectly with the present topological analysis, as does their observation that the vortices are inherently connected to pseudogaps. (In fact, the topological dielectric screening model is even stronger than their model. They assume that the energy for creating vortices in a magnetic field is smaller ("cheap") than in the homogeneous case (no pseudogaps), but at low temperatures the dynamical loops in the present model are actually *native* to the self-organized dopant network [zero, or negative, formation energy].) The data present very serious difficulties for the many other theories they discuss, as those theories are all based on homogeneous mean-field lattice models without specific adaptive dopant configurations. These theories involve different kinds of additional assumptions (including rather far-fetched analogies with elementary particles and liquid crystals [66]), and they do not predict the dopant composition dependence, with its maximum near x = ($x_c$ + $x_0$)/2. The presence of native diamagnetic loops also predicts field-enhanced diamagnetism, with composition dependence parallel to that of the Nernst effect, as observed [67]. Note that both vortex hopping and field loop enhancement assume that there is some "free area" available to vortex dynamics. This "free area" disappears as x → $x_0$, where both effects become smaller.

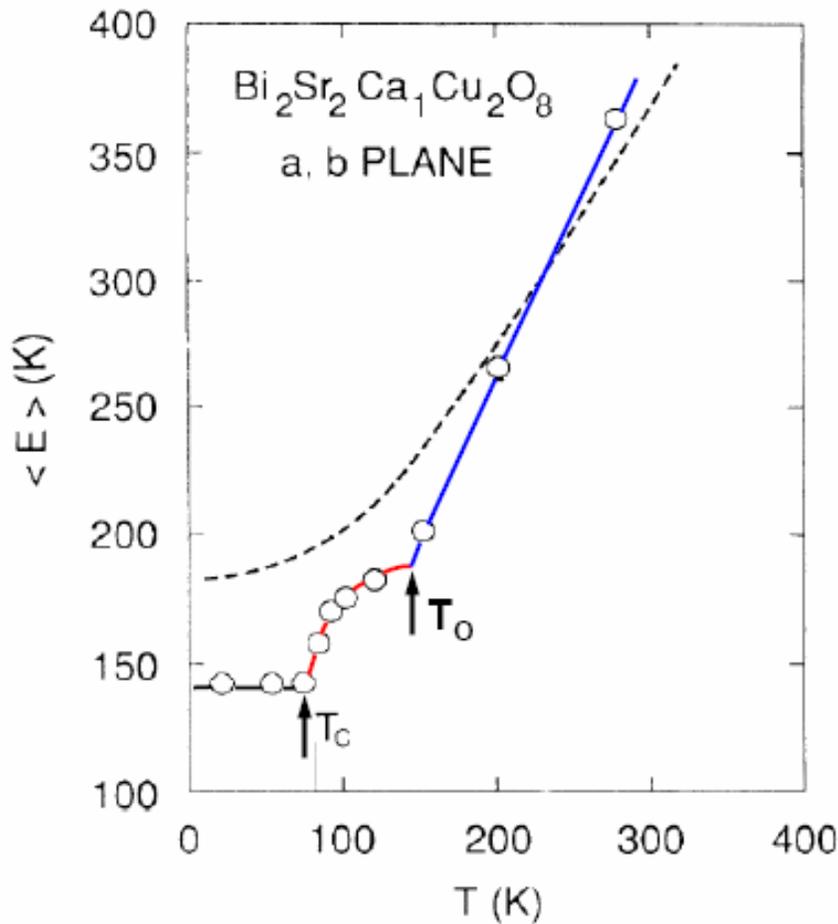

*Fig. 10. <E>(K) of Cu ions [68]. The dashed line is a scaled calculation for a parent (undoped) compound, LCO. Note that doping linearizes and softens <E>(K) for $T > T_0$ = 145 K, and introduces an abrupt stiffening for $T < T_0$, which then shows a "phase transition" as $T \to T_c$.*

If the *B* dopants are indeed in registry with the lattice, one can expect to find evidence in their structures for filamentary formation – providing one is successful in searching the very large (~65,000 papers!) cuprate literature. Several such recent examples have already been noted [13]. However, the most spectacular data found so far are in an old paper [68], which reported data on the average kinetic energy <E>(T) of Cu ions in BSCCO measured by resonant neutron absorption spectroscopy; these are reproduced for the reader's convenience in Fig. 10. The results for optimally doped BSCCO (<R> = 2.14) show a steeper and more *linear* <E>(T) for high T > $T_0$ = 145 K than for undoped LCO (<R> = 2.29); as expected, the smaller <R> softens BSCCO and increases $T_c$. (Thus the onset $T_0$ is about the same for the Nernst effect and Cu kinetic energy arrest. This point could be tested conclusively by measuring $T_0(x)$, which is a very attractive experiment, especially considering that the experiments in [68] were done nearly 20 years ago.) The most interesting behavior, however, is the structural arrest between onset at $T_0$ and $T_c$. This precursive effect becomes very steep near $T_c$, and it can be interpreted as a mean-field transition that "freezes" or merges filamentary end – loop connections. (The strain energy of a filamentary segment is harmonic, and so is proportional to (the length) $l^2$; increasing the length by $\Delta l$ increases the energy by $l\Delta l$, so that $l$ itself is a mean field for filamentary end energies.) Note that the Meissner effect and the onset of superconductivity at $T_c$ is easily explained in the filamentary model in terms of reconnecting filaments or strings to form large loops.

## 11. The "High Energy" Pseudogap

ARPES studies have revealed [69,70] that in addition to the ~ 25 meV pseudogap seen in bulk properties (such as the resistivity, Figs. 3 and 4 of I) there is an incoherent "high energy" pseudogap between $E_1$ = 0.4 eV and $E_2$ = 0.85 eV in the one-electron spectrum. [Aside: energies of order $E_1$, $E_2$ are "high" only in the context of the cuprates. In diatomic semiconductors like GaAs the E(**k**) dispersion relations have been mapped out over an energy range up to 15 eV [53,54] (the bulk plasma energy). One could even speculate that the onset of the incoherent "high energy" range is related inversely to the complexity by the {(number of atoms)/(unit cell)}$^2$.] Nine examples were studied [69,70], in an effort to establish chemical trends, including four samples of BSSCO (2212) ranging from under-to over-doped, with the optimally doped sample both above and below $T_c$. The broad trends were much the same, with small differences in anisotropies in momentum space, which are probably caused by host disorder (for instance, in PbBSCCO [where R(Pb) = 2, as in the perovskite $PbTiO_3$ compared to R(Bi) = 3], or BSCCO(2201) [stacking faults]). In what follows, emphasis is placed on the data on BSCCO (2212) [which has the best-ordered host lattice]. Of course, this gap has nothing to do with spins, superexchange, antiferromagnetism, etc, on lattices. What causes it, and how is it related to the pseudogap at the Fermi energy?

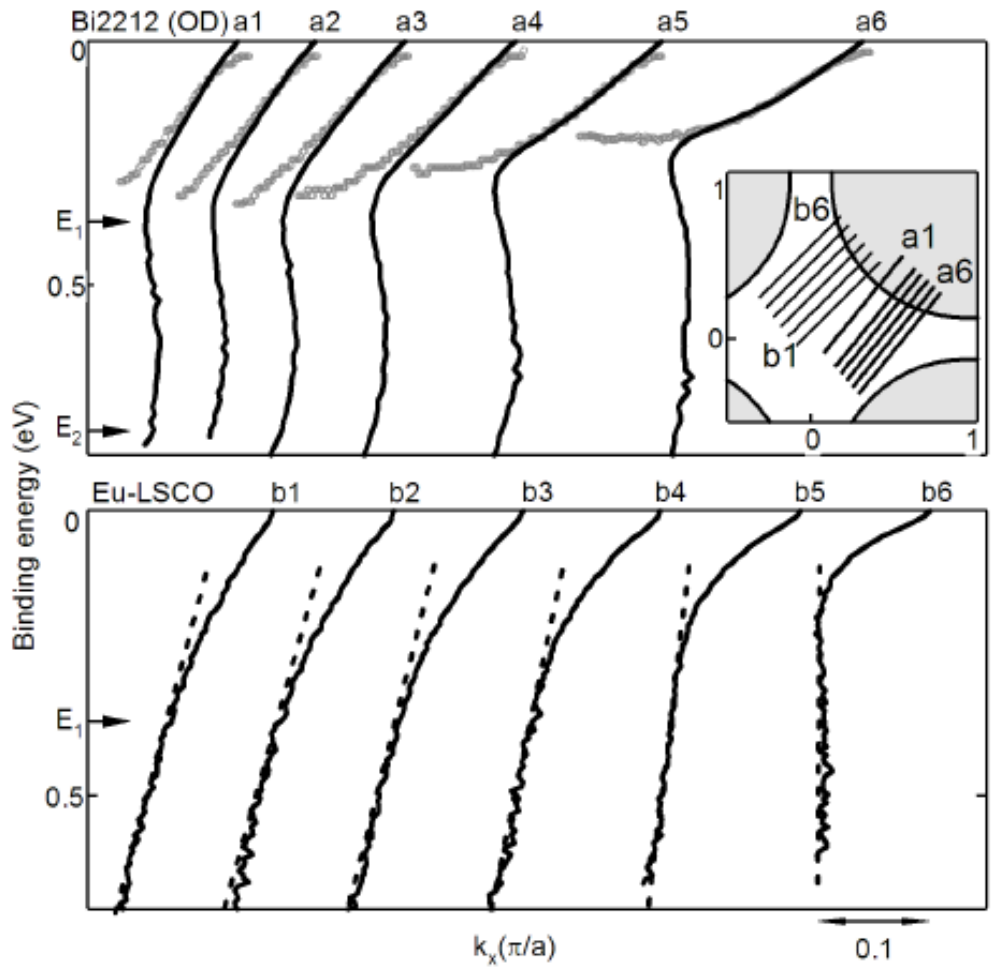

*Fig. 11. Focus on the upper panel, which shows MDC's (solid lines) and EDC's (grey circles) for over-doped Bi(2212). The flattening effects in the EDC's are enhanced in the plots that move from $a_1$ (nearer the nodal (zero gap) direction to $a_6$ (nearer the antinodal (largest gap) direction) [70].*

An example of the BSCCO (2212) data is shown in Fig. 11, reproduced here for the reader's convenience from Fig. 3 of [70]. Between $E_F = 0$ and $E_1 \sim 0.3$ eV the energy distribution curves (EDC) agree well with the momentum distribution curves (MDC), as expected for normal quasiparticles. However, for $E > E_1$ the two curves bifurcate, with the EDC curves flattening out at $E(k) \sim E_1(k_1)$ for $k > k_1$, and the MDC curves becoming nearly vertical at $k = k_1$. It is important to realize such a bifurcation cannot occur in mean field theory, which always yields a single dispersion relation, except near level crossings. Also note that these curves describe peak positions normalized above the elastic background (Fig. 2(a) of [69]). These positions depend on the way that the quasiparticle peaks are incoherently broadened by disorder, especially the disorder generated by dopants, as the mobile quasiparticle states of interest are filaments derived from dopant-centered orbitals. (The broadening of the nodal MDC is shown as an inset to Fig. 2(b) of [69].)

There have been many direct studies of near infrared energy range 0.3 eV – 1.0 eV, including some on "ideal" (micaceous) Bi(2212), yet none of these studies have resolved incoherent anomalies. One obvious advantage of ARPES is that it is able to study separately EDC's and MDC's. Another, less obvious, advantage is that the optical matrix elements $<i|\mathbf{p}|f>$ for photoexcitation to $\sim 20$ eV are very sensitive to glassy disorder. The final states $f$ near 20 eV are very nearly ideal Bloch states, as the dopant potentials perturb such high-energy states only weakly. Thus the optical matrix element projects the amplitude and phase of the glassily disordered initial state $i$ on a periodic state final state $f$, i. e., it samples the periodically coherent part of $i$ in a phase-sensitive way. The same kind of phase coherence is involved in estimating the conductive contribution of a metallic filament to dielectric screening energies. It would be wrong to ignore the phase effects of disorder on the grounds that electron microscopy of cuprates shows apparently ideal lattices, or to attempt to calculate such effects solely from phase-insensitive Debye-Waller factors measured by diffraction. It would be even worse to attempt to describe the effects of disorder on complex orbital motion using a mean-field quasiparticle model.

Worst of all would be a model restricting attention to the $CuO_2$ planes only, and representing the disorder by fluctuations in fictive antiferromagnetic spins at the Cu lattice sites.

With this background one can begin to interpret the main qualitative features of Fig. 11. By constraint counting [1,12], or simply from the constancy of the planar lattice constant, it is clear that the only rigid and strongly periodic planes in the multiplanar cuprates are the isostatically rigid $CuO_2$ planes (R = 2.4). So long as the optically weighted density of quasiparticle states is dominated by initial states *i* in this plane, there will be simple quasiparticles excitations, and there will be good agreement between the EDC and MDC. [The optical weighting varies with energy, and this is why the one-electron self-consistent field calculations of [55] are so successful.] However, for $E > E_1$, the $CuO_2$ planes no longer dominate the optically weighted density of quasiparticle states *i*; instead, other states become important, presumably the SrO semiconductive valence states, where the dopant interstitial O in BSCCO (for example) is thought to reside [60]. Unlike the isostatically rigid $CuO_2$ planes, the SrO layers are soft (or floppy, in the language of constraint theory [1,12]) with R = 2.0, and so they are strongly disordered, both because of their native network instability, and by the presence of coplanar off-lattice oxygen interstitial dopants. Thus the leading effect is a reduction in quasiparticle peak amplitude, or the formation of an incoherent pseudogap for $E > E_1$. It is worth emphasizing that, unlike the pseudogap pinned to $E_F$, this "pseudogap" is not a gap in the density of quasiparticle states *i*, but only in the optically weighted density of quasiparticle states *i* measured by heights and widths of peaks normalized above the elastic background.

What happens to the coherence (relative to that of the $CuO_2$ planes) of states *i* in the SrO planes? There will be a lot of relaxation around the dopant oxygen interstitials, especially near coplanar two-fold coordinated host oxygens. We do not need to know the details of this relaxation, which is quite complex, even in the oxygen neighbors of substitutional Sr in LSCO as studied by EXAFS [13,71]. Our main goal is to understand the mechanism responsible for the bifurcated energy-and momentum-dispersion relations at $E_1$ and $k_1$ shown in the upper panel of Fig. 11. The overall angular picture is that in Bi2212 the ARPES intensity integrated from 0 to 0.8 eV binding energy is largest in the antinodal (10) directions (Fig. 4(b) of [70]), where electron-phonon interactions and both the superconductive and low-energy pseudogap are largest. This suggests that strongly coupled electron-longitudinal optic (LO) phonon interactions increase the coherence of the optical matrix elements; these same LO interactions are responsible for the shifts $\Delta\omega_{LO}$ shown here in Fig. 2(b). The quasiparticles formed by such coupled interactions are called polarons.

There is a large mathematical literature on polarons, but they are usually discussed in the context of simplifying approximations, such as translational invariance, or one LO mode only. In the cuprate context we have spatially inhomogeneous channels formed by nanodomains and layers with different strengths of LO phonon-electron coupling, as well as dopant pinning of carriers. This means that it is not easy to apply polaron formalisms to the cuprates. Broadly speaking, there are two kinds of polarons; weakly coupled polarons are those similar to free carriers, but with small increases in effective mass, while strongly coupled carriers form bound carrier-LO phonon complexes, which give rise to self-trapping. Photoexcited carriers in the cuprates can form bound complexes, providing sufficient phase space is available. In the low (binding) energy region, $E < E_1$, $|k - k_F| < |k_1 - k_F|$, the effective wave lengths for localization are comparable to, or larger than, average channel widths. As E and $|k - k_F|$ increase, at a certain point $k = k_1$ it will become possible for bound complexes to form in channels. This energy $E(k_1)$ will scale with the bare band width, in agreement with experiment [72]. A second factor that should contribute to forming bound complexes is the softening of the host lattice in the (R = 2) SrO layers; as discussed above, increasing E increases the fraction of the quasiparticle density of states in these layers.

Kittel tabulates chemical trends of polaron coupling constants α (sometimes called λ) for the most studied diatomic ionic crystals (alkali and silver halides, where polarons are self-trapped) and a few partially ionic semiconductors [73]. In the former group α ranges from 1.7 (AgBr) to 4.0 (KCl), while in ZnO (a strongly ionic semiconductor, which is a borderline case) $\alpha = 0.85$. SrO, which has the NaCl structure, should have a larger value of $\alpha$ than ZnO, $\alpha(SrO) \sim 1$, and the local value at interstitial O should be much larger, comparable to the Ag halides. Thus polaron self-trapping is expected for interstitial O in the SrO layers of the cuprates. When self-trapping occurs, there is a wide range of momenta for a given polaron energy. Thus formation of bound polarons near O interstitials in the SrO layers explains semiquantitatively the flattening of the EDC's in Fig.11 as E approaches $E_1$, as well as their angular dependence.

What about the MDC's, which show that in the SrO energy range between $E_1$ and $E_2$, the peak momentum $k = k_1$ is apparently constant (energy "waterfall")? This suggests the presence of a charge density wave in the floppy SrO plane generated by the bound polarons and interstitial O's. This CDW is unrelated to the CDW associated with the pseudogap in the $CuO_2$ planes, in accordance with the experimental observation that the high-energy bifurcation is observed with similar strengths in all samples, including overdoped Bi2212, where pseudogaps in the $CuO_2$ planes are much weaker than in underdoped Bi2212. Moreover, although the presence of interstitial O's enhances the softening of the R = 2.0 SrO planes relative to the isostatically rigid (R= 2.4) $CuO_2$ planes, this effect is small, as the SrO planes are floppy anyway [1,12]. Similarly, the SrO dopant centers in LSCO are also floppy.

## 12. Infrared Phase Diagrams

At this point it is convenient to make a few comments on infrared measurements [74]. Relative to ARPES experiments, the information contained in infrared data is limited by several factors (surface quality, exposure to air, convolution of disordered initial with disordered final states). Thus only the phase diagram of the unprocessed reflectivity is easily interpreted. It shows (Fig. 1 of [74]) the largest change between 40K and $T_c$ at around 500 $cm^{-1}$, where R jumps from 0.95 to 0.98. Moreover, the (40K,$T_c$) reflectivity change $\Delta R$ at optimal doping ($T_c = 88K$) is about three times the differences $\Delta R$ observed for slightly Underdoped ($T_c = 66K$) and Overdoped ($T_c = 77K, 67K$), even at 100 $cm^{-1}$.

The effects of inhomogeneities are quite striking: for the Underdoped and Overdoped samples, the (40K,$T_c$) reflectivity changes increase slowly with decreasing frequency, but for the Optimally Doped sample almost all the change $\Delta R$ occurs virtually as a step function around 500 cm$^{-1}$. (In [74] this important point is not mentioned, as the discussion focused on interpreting various features of optical functions derived from deconvolutions of reflectivity and ellipsometric data.) This abrupt appearance of an optical superconductive gap only at optimal doping shows that *at optical wave lengths* the space-filling nature of the optimally doped filamentary network produces a sharp gap distribution, in strong contrast to the STM data [5], which show a broad distribution of superconductive gaps at all doping levels at nm length scales.

The appearance of such homogeneity at a length scale 100 times larger than the nanodomain scale reflects the hierarchical nature of stress relaxation [64]. Probably the simplest explanation for the large step function at optical doping is that only there is the earliest post-cleavage surface oxidation thin film uniform. (Long-term oxidation would probably produce a nearly insulating surface with little superconductivity.) This would also explain why such patchiness was not observed in the STM virgin samples cleaved in vacuum *in situ* [5]. In any event, the disappearance of oxide patchiness at optimal doping is reminiscent of the disappearing isotope effect, also at optimal doping, discussed in the next Section. Both effects reflect space-filling by the filamentary network, with only the oxide patchiness being a surprising consequence of long-range stress fields: it reminds us of the delicate (possibly catalytic) nature initial oxidation kinetics.

Note also that this Optimally Doped jump $\Delta R \sim 0.03$ is still present, but weaker ($\Delta R \sim 0.015$), in the reflectivity data for T = 150K and even T = 220K. The natural interpretation for this is that precursive superconductive pseudogaps – possibly associated with the dopant loops discussed in Sec. 10 (or merely filamentary fragments too short to generate a bulk Meissner effect) are contributing a jump even well above $T_c$. How are these pseudogaps related to the pseudogaps identified in Figs. 3 and 4? To answer this question, we look at the unprocessed reflectivity data for the slightly Underdoped ($T_c$ = 66K) and Overdoped ($T_c$ = 77K, 67K) samples. All these samples exhibit a break in slope for the (40K,$T_c$) curves compared to the (150K, 220K) curves around 1000 cm$^{-1}$. In the Overdoped ($T_c$ = 77K.67K) samples this corresponds to a negative curvature above 1000 cm$^{-1}$, which has been identified as a Fermi liquid lifetime effect [74]; this is what our model predicts, as the Fermi liquid patches can dominate the resistivity even below 1000 cm$^{-1}$, so long as T is above $T_c$. In the underdoped samples the break in slope corresponds to a positive curvature, as the filamentary effects become much more pronounced below $T_c$. These precursive superconductive filamentary pseudogaps are different from the intraplanar interfacial pseudogaps illustrated in Fig. 1. The pseudogaps shown in Figs.3 and 4 could well be influenced by both of these gaps.

It is instructive to compare the optical spectra of the cuprates to those of commercially available semiconductors, which have been analyzed in great detail [53,54]; high quality semiconductor surfaces are easily prepared. The Kramers-Kronig procedures used in [74] to analyze cuprate spectra became standard in the context of semiconductor reflectivity measurements in the 1960's and 1970's for good reasons, namely, the random phase or one-electron approximation is valid for the semiconductors, and the line shapes of interband thresholds, saddle points, etc, are best expressed in terms of $\varepsilon_2(\omega)$, the imaginary part of the dielectric constant, even when excitonic Coulomb interactions deform saddle-point edges [53]. For the ceramic oxides, however, it appears that the main features of the data are seen most clearly in the reflectivity itself.

**13. Isotope Effects**

The rich and impressive data base established for the cuprates over the last 20 years has proved to be more than challenging for various theories, especially the exotic ones that insist that attractive electron-phonon interactions are not responsible for forming Cooper pairs [37,75]. There are large isotope effects in ARPES dispersion relations [27,28] and inhomogeneous gaps measured by STM [26], while recent improvements in ARPES resolution have even led to identification of phonon fine structure in the nodal directions in LSCO [76] similar to the Eliashberg fine structure occasionally seen much earlier in uncontrolled tunneling experiments [77]. However, even in conventional theories many puzzles remain: for instance, why does the isotope effect in $T_c$ become very weak as $T(x)$ approaches either a plateau or $T_c^{max}$ [26]? Some theorists have even argued that the disappearing isotope effect is strong evidence for electron-spin, rather than electron-phonon interactions, as the cause of HTSCD [6,75].

Filamentary theory readily explains the disappearing isotope effect as follows: in underdoped samples, where the isotope shifts are normal, $T_c$ is determined by the Debye phonon energy scale $\omega_D$ for electron-phonon interactions, just as in BCS theory [7]. However, as $T_c(x)$ approaches either a plateau or $T_c^{max}$, the filamentary switching needed to form Meissner or Abrikosov (type II) loops becomes severely restricted: fluctuations in composition x or isotopic weight do not enhance $T_c$, because $dT_c/dx$ is small. This "space-filling" explanation assumes that the larger vibrational amplitudes normally associated with larger $\omega_D$ and smaller masses are hindered by the space-filling which has occurred as T approaches $T_c^{max}$. (With harmonic potentials increasing $\omega_D$ increases $<r^2>$. Most of each filamentary path is planar, so increasing $<r^2>$ is equivalent to increasing x: $T_c(x, \omega_D) = T_c(x/\omega_D)$) Alternatively, the vibrational amplitudes are larger, but this merely increases filamentary overlap, which increases the fraction of Fermi-liquid regions where $T_c = 0$, thus off-setting the enhancement of $T_c$ by the increase in the phonon energy scale $\omega_D$ on filaments. This is the same space-filling mechanism that causes $dT_c/dx$ to go to zero, so the proportionality of the isotope shift and $dlogT_c/dlogx$ is expected, as a natural consequence of the existence of the filamentary intermediate phase between the insulating and Fermi liquid phases, or the reduction in free volume avaialbel for filamentary formation. Once one recognizes the failure of continuum models to describe percolation quantitatively, the "mystery" [6] of the vanishing isotope effect is solved.

The fundamental contrast between planar motion in the isostatic $CuO_2$ planes [1,12] and dopant-assisted tunneling through ultrasoft semiconductive layers like SrO is brought out by site-selective isotope experiments on $T_c$ in YBCO [26]. There the results are quite surprising: the overall trends are described as above, but the cancellation at optimal doping is exact only for oxygen substitution at the apical oxygens or the $CuO_x$ chains (low R sites). At the rigid and almost perfectly crystalline $CuO_2$ planes there is a residual isotope shift even at optimal doping.

## 14. Parallels with Nanocrystalline Transition Metal Oxide Interfaces

The requirement of network self-consistency – especially with respect to internal stress – is very important, and it is responsible for the correlations between HTSC and network glasses shown in Fig. 2(a). However, to obtain more detailed electronic information one should study simple systems that are closer to the cuprates, such as the nearly planar-lattice-constant-matched cubic perovksites $SrTiO_3$ and $LaAlO_3$ heterointerfaces [78] and thin-film nanocrystalline $TiO_2$, $ZrO_2$, and $HfO_2$ [79]. There are many striking parallels between the results obtained both electrically and spectroscopically (probing both valence and conduction band energies separately) in these materials and in the cuprates. The most striking similarity is that with increasing grain size electronic band coherence appears at a grain size of 3 nm, the same size as the average gap nanodomain seen in the cuprates by STM [5]. This validates the fundamental assumption of the filamentary percolative model, that the "ideal" metallic $CuO_2$ planes of the cuprates are in fact broken up into coherent 3 nm units separated by interfaces with pseudogaps (see Fig. 1, which dates from 1990!).

## 14. Conclusions

Filamentary theory, based on topological reasoning, is well-suited to discussing strongly disordered oxides like the cuprates. These materials are all anomalously soft, with R < 2.4 (Fig. 2 of I). The architectonics of such anomalously soft (glassy or nearly glassy) oxides is quantitatively described by constraint theory. There are at least *five* (!) space-filling features of the huge 65,000 paper HTSC data base that are inexplicable with analytic models: the Ando II line (Fig. 4 of I); the $\Delta Z$ jumps (Fig. 5 of II); the exciton relaxation time jump (Fig. 7 of II); the reflectivity jump (Fig. 1 of [74]), and the near-disappearance of the isotope effect, all occurring near optimal doping. While the theory is far from complete, it is realistic because it takes into account the strongly disordered, off-lattice glassy character of doped cuprates, which is by now extremely well documented by STM experiments, especially those of Davis *et al* [5]. Lattice models are fixed, ignore this disorder, and generally are incapable of recognizing or explaining space-filling features. They cannot expect to be any more successful in describing the cuprates than spin-glass lattice models are in discussing the composition-dependent properties of real glasses like window glass. Moreover, plane wave models cannot consistently explain the observed d-wave anisotropy of the energy gap and the large magnitudes of $T_c$ [80], or the $T_c$ of $YBCO_{6.5}$ if all the O ions are fixed at ideal lattice sites [81]. Strong disorder and space-filling complicate the theory, but they also provide the flexibility and adaptability that make HTSC possible [1].

*Postscript.* There are spatial inhomogeneities in manganite perovskites (for instance, $(La,Ca)MnO_3$) similar to those in the cuprates. Recent large-scale 2-dimensional Monte Carlo calculations using the travelling cluster approximation show [82] that, even with exchange interactions, on-lattice polaronic carrier charge inhomogeneities also have a filamentary structure (similar to that found in classical models [52]), even though the carriers are not pinned to the (immobile Ca) dopants. Fourier analysis of STM experiments on BSCCO has shown [83] that dopant (oxygen interstitials!)-phonon coupling is very large in the normal state, and even larger in the superconductive state, as assumed here in Section 6 to explain the evolution of ARPES Fermi arcs. The Fourier analysis does not include filamentary effects, which are necessary to understand the origin of the interactions they parameterize.

Sec.11 discussed the "high energy waterfall" in BSCCO and related compounds, and explained the results primarily in terms of self-trapped polarons centered near interstitial oxygen dopants. The waterfall has also been studied [84] in optimally doped LSCO, with results that are qualitatively different from those obtained on the BSCCO family [69,70]. In LSCO the Sr dopants are nearly substitutional, and the connectivity of the internal dopant network is determined by the large displacements of the oxygens that are first (10) and second (21) neighbors of the Sr dopants [71]. These displacements are

21strongly influenced by the host lattice structure and the relative (10) and (21) weights. Thus the self-trapping effects can be expected to exhibit some kind of d-wave anisotropy, as observed: a remarkable effect, and very pretty [84]. Given that the (10) first neighbor interactions must be much larger than the (21) second neighbor interactions, the form of the d-wave anisotropy is explained, as well as its large magnitude ($E_1(\pi/4) \sim 10E_1(\sim 0)$). The most surprising feature of these data is the small value of $E_1(\sim 0)$. It seems likely that the onset of decoherence at such low energies in LSCO (relative to the micaceous BSCCO results) is caused by surface roughening. The dopant dependence from $x \sim 0$ up to $x \sim 0.3$ of the largest $E_1(\pi/4)$ is nearly linear in x. This could well reflect a decrease in surface roughening as dopants are added, making the surface softer and less brittle.